\documentclass{aa}
\usepackage[varg]{txfonts}

\usepackage{amsfonts}       %
\usepackage{graphicx}
\usepackage{natbib}
\usepackage{textcomp}
\usepackage{xcolor}
\usepackage{hyperref}
\usepackage{comment}
\usepackage[nolist]{acronym}
\usepackage{ulem}

\usepackage{soul}

\newcommand{\popratio}[3]{\ensuremath{ \delta_{#1,#2}^{\mathrm{#3}}(z) = \frac{n_{\mathrm{^{#1}Ca,\,#3}}(z)}{n_{\mathrm{\,^{#2}Ca,\,#3}}(z)}}}

\newcommand{\popratioshort}[3]{\ensuremath{\delta_{#1,#2}^{\mathrm{#3}}}}

\newcommand*{\ie}{i.e.\@\xspace}
\newcommand*{\eg}{e.g.\@\xspace}

\newcommand{\Mm}{~\mathrm{Mm}}
\newcommand{\G}{~\mathrm{G}}

\newcommand{\km}{~\mathrm{km}}

\newcommand{\tauone}{~\ensuremath{\tau_{500}=1}}
\newcommand{\kms}{\ensuremath{\, \mathrm{km\,s^{-1}} }\xspace}

\newcommand{\caisotopefourty}{\ensuremath{^{40}\mathrm{Ca}}\xspace}

\newcommand{\mghk}{\ion{Mg}{II} h\&k\xspace}

\newcommand{\mgk}{\ion{Mg}{II} k\xspace}

\newcommand{\cahk}{\ion{Ca}{II} H\&K\xspace}

\newcommand{\snapfourNineNine}{\texttt{muram\_en\_499000\_379s}\xspace}

\newcommand{\snapfiveOneEight}{\texttt{muram\_en\_518000\_503s}\xspace}

\newcommand{\caefft}{\ion{Ca}{ii}~$\lambda854.2$~nm\xspace}

\title{\caefft in an enhanced network region simulated with MURaM-ChE}

\titlerunning{\caefft in an EN region simulated with MURaM-ChE}
\newcommand{\muram}{MURaM\xspace}

\newcommand{\intensity}{\mathrm{\,n\, J\, m^{-2} \, s^{-1}\, Hz^{-1}\, sr^{-1}}}

\defcitealias{2024A&A...692A...6O}{Pub~I}

\author{ P.A. Ondratschek\inst{1}, D. Przybylski\inst{1}, H.N. Smitha\inst{1}, R.H. Cameron\inst{1}, S.K. Solanki\inst{1}}
\authorrunning{Ondratschek et al.}
\titlerunning{\caefft in MURaM-ChE EN}

\institute{Max Planck Institute for Solar System Research,Justus-von-Liebig-Weg 3, 37077 G\"{o}ttingen, Germany\\
\email{ondratschek@mps.mpg.de}}
\begin{document}

\begin{acronym}

\acro{lte}[LTE]{local-thermodynamic-equilibrium}
\acro{rte}[RTE]{radiative transfer equation}
\acro{te}[TE]{thermodynamic equilibrium}
\acro{ne}[NE]{nonequilibrium}

\acro{nlte}[NLTE]{non-local-thermodynamic-equilibrium}
\acro{se}[SE]{statistical equilibrium}
\acro{los}[LOS]{line-of-sight}
\acro{eos}[EoS]{equation of state}

\acro{crd}[CRD]{complete frequency redistribution}
\acro{prd}[PRD]{partial frequency redistribution}
\acro{rh15d}[RH1.5D]{Rybicki \& Hummer 1.5D RT code}
\acro{rt}[RT]{radiative transfer}
\acro{mhd}[MHD]{magnetohydrodynamics}
\acro{rmhd}[rMHD]{radiative-magnetohydrodynamics}
\acro{iris}[IRIS]{Interface Region Imaging Spectrometer }
\acro{muram}[MURaM]{Max Planck Institute for Solar System Research/University of Chicago Radiation Magneto-hydrodynamics}
\acro{muramche}[MURaM-ChE]{chromospheric extension of MURaM}
\acro{clv}[CLV]{center-to-limb variation}
\acro{euv}[EUV]{extreme ultra violet}
\acro{nir}[NIR]{near-infrared}
\acro{nuv}[NUV]{near-ultraviolet}
\acro{uv}[UV]{ultraviolet}
\acro{ali}[ALI]{approximate Lambda iteration}
\acro{ff}[ff]{free-free}
\acro{qs}[QS]{quiet Sun}
\acro{soup}[SOUP]{Solar Optical Universal Polimeter}
\acro{sst}[SST]{Swedish $1 \, \mathrm{m}$ Solar Telescope}
\acro{chromis}[CHROMIS]{CHROMospheric Imaging Spectrometer}
\acro{gong}[GONG]{Global Oscillation Network Group}
\acro{dot}[DOT]{Dutch open telescope}
\acro{ar}[AR]{active regions}
\acro{fb}[fb]{free-bound}
\acro{bf}[bf]{bound-free}
\acro{en}[EN]{enhanced network}
\acro{fov}[FOV]{field-of-view}

\acro{bb}[bb]{bound-bound}
\acro{fwhm}[FWHM]{full width at half maximum}
\acro{ssd}[SSD]{smale-scale dynamo}
\acro{fts}[FTS]{Fourier Transform Spectrograph}

\acro{im}[IM]{isotope model}
\acro{sim}[SIM]{single isotope model}
\acro{cm}[CM]{composite model}
\acro{ftsatlas}[Hamburg FTS atlas]{Hamburg Fourier-transform-spectrograph atlas}

\acro{sfp}[SFPs]{strong-field-profiles}
\acro{wfp}[WFPs]{weak-field-profiles}

\acro{ufp}[UFPs]{upflow-profiles}

\acro{dfp}[DFPs]{downflow-profiles}

\end{acronym}

\abstract{The \caefft line is widely used to study the chromosphere of the Sun. In the quiet Sun, the spatially averaged line profile shows a red asymmetry and a redshift of the line center. It is known that the effect of isotopic splitting must be taken into account in the forward modeling to reproduce the observed asymmetry. So far, no numerical model could match an average observed line profile in terms of the line width and asymmetry.}
{Our goal is to investigate how well a simulation computed with the chromospheric extension of the MURaM code (MURaM-ChE) reproduces the spatially averaged \caefft line profile. We aim to determine the contributions from the isotopic splitting versus the dynamics in the atmosphere to the resulting line width and asymmetry. For this purpose, we forward model the line based on a simulated enhanced network region.}
{Our study builds on forward modeling of the \caefft line in a series of MURaM-ChE simulation snapshots representing an enhanced network region. We solve the radiative transfer problem three times, once considering only the most abundant isotope of calcium in the atmosphere, once taking six calcium isotopes into account, and finally using a single ''composite‘‘ atom model, which mimics the presence of all six isotopes.}
{We find the forward modeled spatially and temporally averaged spectra to be in good agreement with the \ac{ftsatlas} observation of the quiet Sun. In order to match the observed line width, the simulated atmosphere must be sufficiently dynamic. The typical red-asymmetry can only be reproduced by taking the isotopic splitting effect into account, as suggested in the literature. The closer match between the new model and the observations compared to earlier numerical models is a result of the higher rms-velocity in the MURaM-ChE chromosphere. The center of the spatially averaged line profile tends to be slightly red-shifted, which is a result of a net downflow velocity at the formation height of the line center intensity. This does however not imply average mass downflow. We find the  composite atom model is a good approximation to the full isotope computation, but shows some differences in the line core and asymmetry.
}
{We show that forward modeling of the \caefft line from an MURaM-ChE simulation can result in a close match to the line shape of an average quiet Sun observation. The atmosphere must be sufficiently dynamic to match observed line width.  Our results confirm that it is important to include the isotopic splitting effect of calcium in modeling the \caefft line. }

\keywords{magnetohydrodynamics, radiative transfer, Sun:chromosphere}
\maketitle

\section{Introduction}
\label{sec:introduction}
The chromosphere is a component of the solar atmosphere that connects the photosphere to the overlying hot corona. 
Spectral line formation in the chromosphere is complicated, and the interpretation of the observed fine structure is difficult \citep[see \eg,][]{2010MmSAI..81..565R, 2019ARA&A..57..189C}. Observations of strong spectral lines that form in the chromosphere are the key to inferring thermodynamic quantities such as velocity and temperature, as well as the magnetic field. Interpreting the observations, however, requires a detailed understanding of line formation. The \ac{ne} and \ac{nlte} conditions in the chromosphere are reflected in the line formation. Among the available chromospheric diagnostics, the \ion{Ca}{ii} infrared triplet lines have advantageous properties. Their location in the infrared makes them observable from the ground. They are not subject to scattering as much as other chromospheric lines such as \cahk and H$\alpha$ \citep{1989A&A...213..360U} and to \ac{ne} effects \citep{2011A&A...528A...1W}, which makes them easier to model and interpret. In addition, the \ion{Ca}{ii} infrared triplet lines, in particular \caefft, are good candidates for magnetic field inference in the chromosphere through inversions \citep[\eg,][]{2012A&A...543A..34D,2016MNRAS.459.3363Q,2016ApJ...826L..10S,2017SSRv..210..109D}.

Spatially averaged observations of the \ac{qs}, for example with the \ac{ftsatlas} \citep[][]{1984SoPh...90..205N}, indicate a red-asymmetry in the core of the \caefft line \citep{2006ApJ...639..516U}, which is in contrast to photospheric lines that show a blue-asymmetry \citep[see \eg,][]{ 2000A&A...359..729A,2005oasp.book.....G}. In the photosphere, a different area coverage, line strength, and composition of upflows in the granules and downflows in the intergranular lanes lead to the observed blue asymmetry. The asymmetry of spectral lines is typically inferred from the line bisector. In the case of the \caefft line, the bisector has the shape of a mirrored ''C‘‘ letter and is, therefore, often called an ''inverse C-shape bisector‘‘. In contrast, other chromospheric lines such as H$\alpha$ show a C-shaped bisector \citep{2013SoPh..288...89C}.

\citet{2006ApJ...639..516U} suggested that the inverse C-shape is a result of acoustic waves that steepen into shocks at chromospheric heights, similar to the analysis of \ion{Ca}{ii} K2V bright grains \citep{1997ApJ...481..500C}. The wave propagation results in an asymmetry in the time the plasma spends in up- and downflows. By using three-dimensional convective simulations from \citet{2000A&A...359..729A}, \citet{2006ApJ...639..516U} found a red asymmetry in the spatially averaged \caefft profile, but of only roughly one-fifth of the observed amplitude. Numerical models of the solar chromosphere, in combination with three-dimensional \ac{rt} computations, failed to reproduce the observed asymmetry \citep[\eg,][]{2009ApJ...694L.128L}. The profiles instead showed symmetric line cores.

\citet{2014ApJ...784L..17L} included all isotopes of calcium that exist stably in the solar chromosphere in the \ac{rt} calculation. By solving the \ac{rt} problem simultaneously for six isotopes of calcium, the authors found significant differences compared to the computations where only the most abundant isotope was used. For their analysis, the authors used the FAL C model \citep{1993ApJ...406..319F}, a time series from a RADYN simulation \citep{1992ApJ...397L..59C}, and a model computed with the Bifrost code \citep{2011A&A...531A.154G}. The synthesized \caefft profiles from all three model atmospheres resulted in a red asymmetry in the line core, that is an inverse C-shaped bisector. 

While there has been progress in explaining the observed \caefft line, the forward modeled spectra still show discrepancies. Although the spectra show a red asymmetry, the exact shapes do not match the observations, and the bisector amplitude is too low. For chromospheric lines, the spatially averaged line width of forward-modelled spectra is typically too low  \citep[see \eg,][]{2016ApJ...826L..10S, 2018A&A...619A..60J,2023ApJ...944..131H}. The mismatch of the line width is assumed to be due to a too-low magnitude of dynamical motions \citep[see \eg,][]{2016A&A...585A...4C}.

In recent work, \citet[][hereafter \citetalias{2024A&A...692A...6O}]{2024A&A...692A...6O} showed that new models of the chromosphere are able to match the line properties of the \mghk lines relatively closely. The authors find that high maximum velocity differences in the chromosphere may explain the observed line width. The atmosphere model resembles an enhanced network region similar to the setup of the public Bifrost snapshot \citep{2016A&A...585A...4C}. In this work, we study the spatially averaged \caefft line in the same model of the chromosphere as in \citetalias{2024A&A...692A...6O}, which is simulated with the recently developed chromospheric extension of MURaM \citep[\acs{muramche}\footnote{Max Planck Institute for Solar System Research/University of
Chicago Radiation Magneto-hydrodynamics with a chromospheric extension.},][]{2022A&A...664A..91P}. We aim to identify the requirements to match the observed spatially averaged line profile. While the effect of isotopic splitting on the line asymmetry was demonstrated by \citet{2014ApJ...784L..17L}, it is not clear how important this effect is in the \acs{muramche} model used here. We therefore study the role of the dynamic motions versus the isotopic splitting effect in the resulting spatially averaged line profile.

This paper is structured as follows. In Sect. \ref{sec:methods_atmosphere_and_rt} we present the atmosphere model and \ac{rt} calculations. In Sect. \ref{sec:results} we present our results. In Sect.~\ref{sec:discussion} we summarize and discuss the results, and in Sect.~\ref{sec:conclusions} conclusions are given.

\section{Model atmosphere and forward modeling}
\label{sec:methods_atmosphere_and_rt}

We model the \caefft spectral line in an atmosphere calculated with \acs{muramche}, which is a \ac{rmhd} code. The \muram code originally included the physics required to simulate near-surface convection \citep{2005A&A...429..335V} but was limited to a \ac{lte} approximation of the radiative and atomic physics. \citet{2017ApJ...834...10R} extended the code to include optically thin losses and heat conduction along magnetic field lines to model the corona. 

Recently, \citet{2022A&A...664A..91P} developed the chromospheric extension of MURaM, which includes an \ac{ne} treatment of hydrogen in and above the photosphere. The convection zone is modeled by a non-ideal \ac{eos} generated with the free-EoS package \citep{2012ascl.soft11002I}. Above the photosphere, a \ac{ne} \ac{eos} is used, which includes a time-dependent treatment of hydrogen ionisation following the prescriptions in \citet{1999MsT..........1S, 2006A&A...460..301L, 2007A&A...473..625L}. In the \ac{eos}, all non-hydrogen elements are treated in \ac{lte}. The code includes chromospheric line losses and optically thin losses in the corona, as described in \cite{2012A&A...539A..39C}. In addition, the code includes 3D \ac{euv} back-heating of the chromosphere similar to \cite{2012A&A...539A..39C}. Radiative losses in the photosphere and low chromosphere are calculated using a four-band multigroup short-characteristics \ac{rt} scheme \citep{1982A&A...107....1N,2004A&A...421..741V}, which was extended to include scattering effects as in \citet{2000ApJ...536..465S} and \citet{2010A&A...517A..49H}. The solar abundances are taken from \citet{2009ARA&A..47..481A}.
The formation and dissociation of $H_2$ molecules are treated in \ac{ne}, while $H_2^{+}$ and $H^{-}$ are treated in chemical equilibrium. A slope-limited diffusion scheme is included as described in \cite{2014ApJ...789..132R} and \citet{2017ApJ...834...10R}. 

The simulation setup resembles an enhanced network region, similar to the public Bifrost snapshot, and is the same as the one used in \citetalias{2024A&A...692A...6O}. The simulation domain spans $24 \Mm\times 24\Mm \times 24 \Mm$ with a resolution of $23.46 \km$ horizontally and $20 \km$ vertically. The convection zone extends roughly $7 \Mm$ below the \tauone~ surface, and the atmosphere covers approximately $17 \Mm$ in height. We use the convention that a positive vertical component of the velocity corresponds to an upflow in the atmosphere. 

Details about the se up of the simulation can be found in \citet{2022A&A...664A..91P} and \citetalias{2024A&A...692A...6O}. Here, we give a brief summary of how the system reached the state in which it was used for the forward modeling. The simulation was initially set up as an \ac{ssd} simulation with a horizontal extent of $12 \Mm \times 12 \Mm$ \citep{2025A&A...703A.148P}. The horizontal domain was extended by tiling it $2 \times 2$ and running it for another $8$ hours of simulation time to break the periodicity. After that, a bipolar magnetic feature with roughly $8 \Mm $ separation between the poles was added, similar to the enhanced network model of the public Bifrost simulation \citep{2016A&A...585A...4C}. The simulation was then run for another $1.5$ hours after which the \ac{ne} treatment of hydrogen was turned on. Following this, the simulation was run for another $10 \min$ to let the populations settle into a new equilibrium. We then computed a time series of $10 \min$, which we use for our analysis. The simulation is overall in a steady state but shows signatures of the typical $5 \min$ oscillations in the photosphere and $3$--$5 \min$  oscillations in the chromosphere. For a detailed analysis of the oscillations, a longer time series is required.

We use the RH1.5D code \citep{2001ApJ...557..389U, 2015A&A...574A...3P} to synthesize the \caefft spectral line. In this code, each vertical column in the simulated atmosphere is treated independently as a plane-parallel atmosphere. This is also called the 1.5D \ac{rt} approach. We perform the spectral line synthesis at a heliocentric angle $\theta$ such that $\mu=\cos(\theta)=1$ corresponding to a disk center observation. By doing so, the \ac{los} components of vector quantities such as velocity or magnetic field correspond to the vertical component in the simulation grid.

We perform three sets of spectral synthesis. Firstly, following \citet{2014ApJ...784L..17L} we treat the all six stable isotopes of calcium, $^{40}$Ca,$^{42}$Ca,$^{43}$Ca,$^{44}$Ca,$^{46}$Ca,$^{48}$Ca as separate atoms that are all treated in \ac{nlte} in the \ac{rt} computation, hereafter \ac{im}. 
For the \ac{rt} computations five-level-plus-continuum model atoms of \ion{Ca}{ii} were constructed similar to  \cite{2014ApJ...784L..17L} based on the experimental data from \cite{ 1992PhRvA..45.4675M} and \citet{1998EPJD....2...33N}.

Secondly, we used a \ac{cm}. This is an approximation to the first computation with the advantage that only one atom needs to be computed in \ac{nlte} to solve the \ac{rte}. The effect of multiple isotopes is mimicked by modifying the absorption coefficient of the model atom. This is achieved by summing up Voigt profiles centered at the rest wavelengths of the corresponding isotopes. The sum is weighted by the relative abundance of the isotopes.

Finally, we used a five-level-plus-continuum atom model where it is assumed that the most abundant isotope in the solar atmosphere (\ie, $^{40}$Ca) is the only present calcium isotope in the atmosphere, hereafter \ac{sim}. By doing so, we can test the effects of multiple isotopes on the line width and asymmetry by comparing it to the computation where only the most abundant isotope was considered. The computations were conducted in the approximation of \ac{crd}, which is sufficient for the \caefft spectral line \citep{1989A&A...213..360U}.

\section{Results} 
\label{sec:results}
 \begin{figure*}
   \centering
   \includegraphics[width=\textwidth,clip]{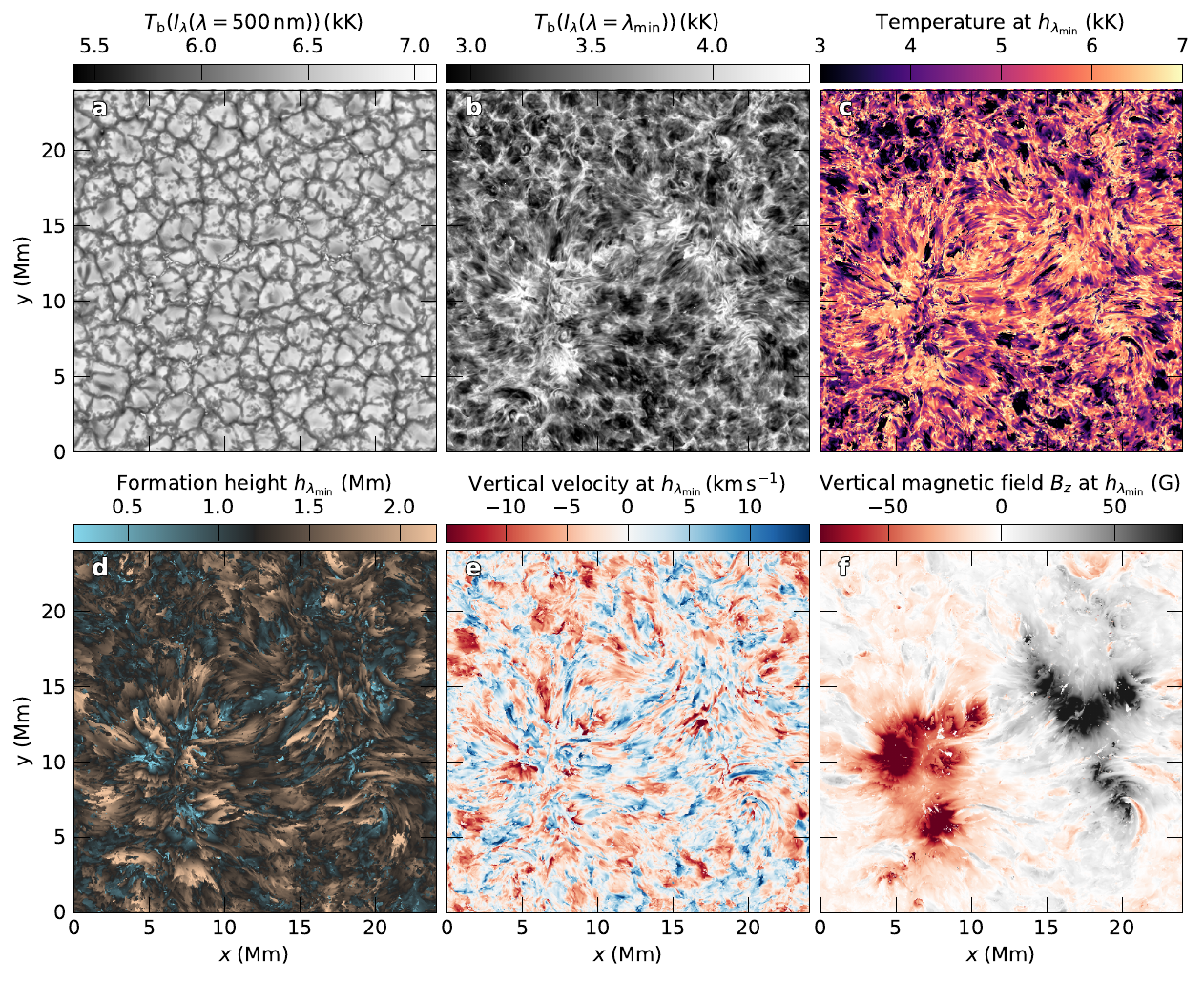}
      \caption{Overview of intensity, and atmospheric properties at the formation height of the intensity profile minimum. Panel (a) shows the continuum intensity at $\lambda=500\, \mathrm{nm}$ as brightness temperature, panel (b) shows the intensity at the profile minimum $\lambda_{\mathrm{min}}$ of the \caefft spectral line as brightness temperature, panel (c) shows the temperature at the formation height of the line profile minimum, panel (d) shows the formation height of the line profile minimum $h_{\lambda_{\mathrm{min}}}=h(\tau_{\lambda,\mathrm{min}}=1)$, panel (e) shows the vertical velocity at $h_{\lambda_{\mathrm{min}}}$, and panel (f) shows the vertical component of the magnetic field at $h_{\lambda_{\mathrm{min}}}$. The line profile minimum and the corresponding formation height are individually determined for each column in the atmosphere. The color scale limits are clipped to increase the contrast of the images (see text). The intensity (panels a and b), as well as the formation heights that were used to create panels (c, d, e, and f), originate from the \ac{im} \ac{rt} computation from the snapshot \snapfiveOneEight.}
         \label{fig:fig2_int_vz_bz_temp}
\end{figure*}   

We present our results as follows: first, we describe the intensity images at $500 \, \mathrm{nm}$ and at the line profile minimum, together with the temperature, the vertical component of the velocity, and the vertical component of the magnetic field at the corresponding formation height of the line profile minimum. We then show the spatially averaged spectra with corresponding line bisectors based on the three sets of \ac{rt} calculations. Following this, we analyze how the spatially averaged spectrum and its asymmetry depend on different regions in the atmosphere that are selected by the local vertical magnetic field and vertical velocity in each column.

\subsection{Intensity and atmospheric properties at the formation height of the line core}
\label{sec:results_intensity_map_and_atmosphere}

In the here presented model, the \caefft line profile minimum forms approximately between $0.5 \Mm$ and $2.5 \Mm$, where $z=0 \Mm$ is the average height of the $\tau_{500}=1$ surface. Within these heights, the individual profiles form under a wide range of physical conditions. We therefore present in Fig.~\ref{fig:fig2_int_vz_bz_temp} an overview of the continuum intensity at $\lambda= 500 \, \mathrm{nm}$, the line profile minimum intensity $I_{\mathrm{min}}$, and formation height at the wavelength position of the minimum intensity. We determine the formation height by first finding the wavelength of the intensity minimum $\lambda_{\mathrm{min}}$ at each single spectrum, and then locating the height position $h$ in the atmosphere where the optical depth at this wavelength reaches unity, that is $h(\tau_{\lambda,\mathrm{min}})=1$. The wavelength position of $\lambda_{\mathrm{min}}$ is obtained in a window of $\pm 0.5 \, \AA$ around the rest wavelength of the most abundant calcium isotope $^{40}$Ca. We note that the synthetic profiles in the line core can be complex with additional emission peaks. A determination of the line profile minimum in the line core is thus not always unique. In addition, we show temperature, vertical velocity, and vertical magnetic field at the formation height of the minimum intensity.

In the continuum intensity map (Fig.~\ref{fig:fig2_int_vz_bz_temp}, panel a), the granulation patterns at the bottom of the photosphere are visible. Small bright structures reveal the presence of the network magnetic field in the simulation domain. The intensity map of the line-core minimum (panel b) shows web-like structures throughout the whole simulation domain, except above the strong network fields, where larger bright structures are visible. The web-like structure is a result of shock waves in the lower atmosphere that expand horizontally with height and interfere with each other. By compressional (adiabatic) heating, the temperature (panel c) is locally increased, leading to the visible pattern \citep[see \eg,][]{2004A&A...414.1121W}.

The temperature map (panel c) underlines the correlation between the line core intensity and shock patterns in the quiet regions at roughly $0\leq y / \Mm\leq 5$ and $17\leq y / 
\Mm\leq 24$.
Above the strong network fields at roughly $5\leq y / \Mm\leq 17$, the temperature is higher and is often associated with lower formation heights (panel d). Small, bright pixels can be seen as the formation height can change sharply in 1.5D \ac{rt}. In addition, some intensity profiles show core reversals with multiple peaks, such that a clear definition of the line profile minimum is not always straightforward.

The formation height (panel d) reveals the corrugated surface over which the \caefft line core forms in the \acs{muramche} model. In the horizontal center of the simulation domain, the formation height indicates loop-like regions forming in the mid-to-upper chromosphere and shows fine structure. The fibrilar structures are not visible in the intensity image (panel b), presumably because of missing horizontal \ac{rt}. A similar effect was found for the H$\alpha$ line \citep[][]{2012ApJ...749..136L} and the \mghk lines \citep{2017A&A...597A..46S} in the Bifrost public snapshot.

The vertical velocity (panel e) indicates where the line core forms in upflows (blue) or downflows (red). Qualitatively, it can be seen that structures associated with upflows are more concentrated than downflows. The area coverage of rays, where the line core forms in a downflow, is $57\%$ vs. $43\%$ in an upflow. The average upflow velocities are $3.42 \kms$ vs. $-3.84 \kms$ in downflows. Both upflows and downflows reach maximum values of approximately $20 \kms$ magnitude. The higher magnitude of the downflow velocities results from lower densities \citep[see also][]{2006ApJ...639..516U}, which also agree with the larger area (and volume) coverage of downflows in the chromosphere.

Finally, in panel (e), we show the vertical component of the magnetic field at the formation height of the line core. The network fields expand in the chromosphere \citepalias[compared with the photosphere; see][Fig. 1a]{2024A&A...692A...6O}, forming a canopy. The bipolar magnetic field patterns roughly coincide with the increased intensity pattern in panel (b). Small irregular features such as at $(x,y)\approx (8\, \Mm, 12.5\, \Mm)$ result from ambiguities in the determination of the line profile minimum due to complex line core profile shapes.

\subsection{Spatially averaged line profiles}
\label{sec:results_average_profiles}

 \begin{figure*}
   \centering
   \includegraphics[width=\textwidth,clip]{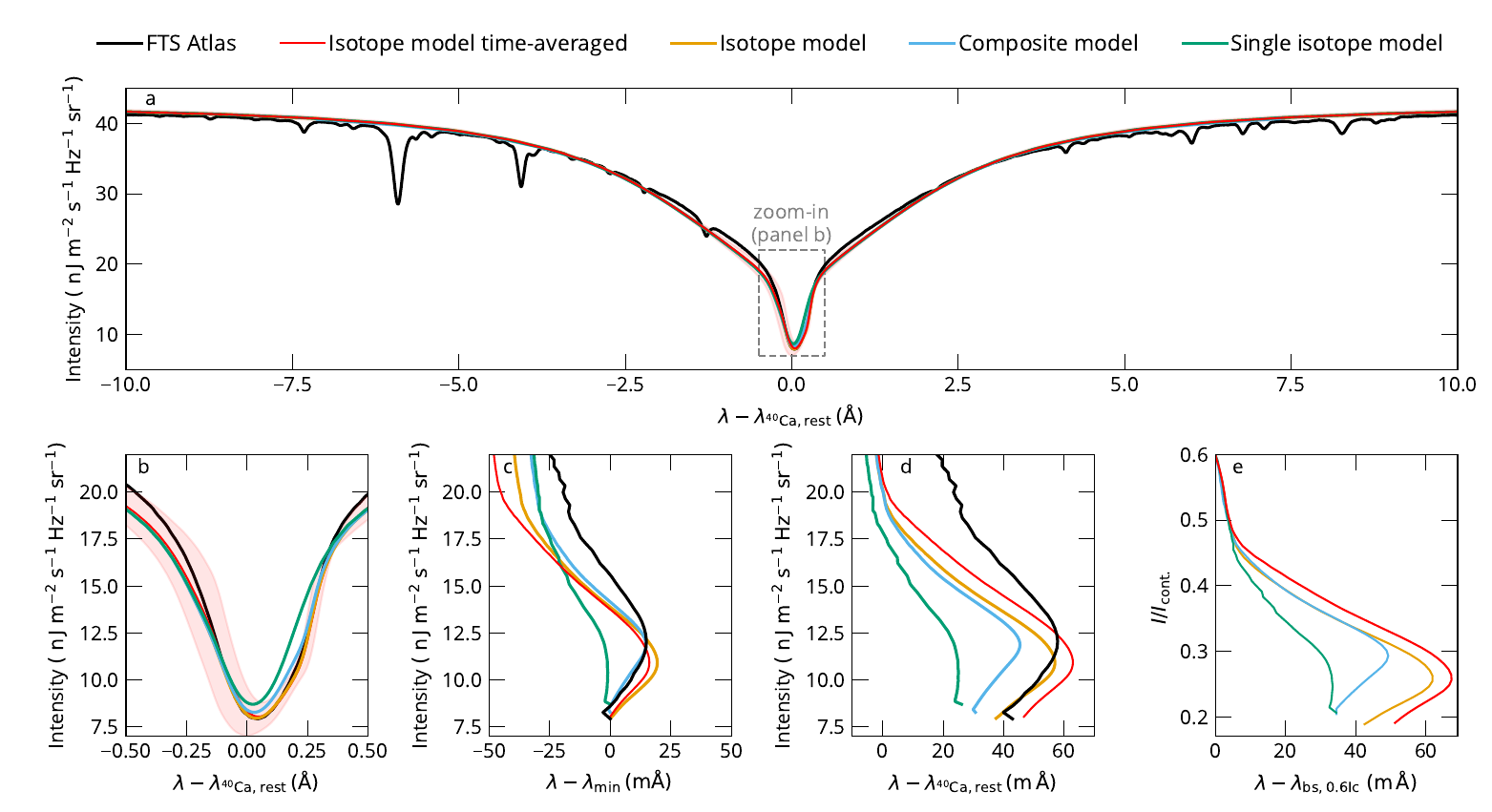}
      \caption{Comparison of spatially averaged profiles (panels a and b) and corresponding bisectors (panels c, d, and e) of the \caefft line. The black curves show spatially averaged \ac{qs} spectra from the \ac{ftsatlas}. The red curve shows a time average of the full isotope computation of four snapshots separated by $2\,\mathrm{min}$ (see Appendix \ref{app:time-variation} for more details) together with the standard deviation multiplied by a factor of eight in light red color (not visible in panel a but in panel b). The orange, green, and blue curves show data from the single snapshot  \snapfiveOneEight. In orange, we show the spatially averaged synthetic spectra computed by taking all isotopes of calcium into account. The other curves show similar profiles but computed without isotopic splitting (green), and using a composite model atom (blue). 
      The wavelength axis of the average profiles is centered at $\lambda - \lambda_{\mathrm{^{40}Ca,rest}}$ where $\lambda_{\mathrm{^{40}Ca,rest}}$ is the rest wavelength of $^{40}\mathrm{Ca}$, the most abundant calcium isotope. The bisectors are presented in three different ways. Once they are all centered on the wavelength of the corresponding line profile minimum (panel c), once the bisectors are shown on an absolute wavelength scale (panel d), and once the bisectors are centered on the wavelength where the bisector intensity is $60\%$ of the continuum intensity (panel e). 
      }
         \label{fig:fig1_av_spectra_and_bisector}
\end{figure*}   

In the following, we present the spatially and temporally averaged \caefft spectral line profile calculated from a series of four snapshots, where we included all six isotopes of calcium (\ac{im}, see Sect.~\ref{sec:methods_atmosphere_and_rt}). For one instance in time, snapshot \snapfiveOneEight\footnote{We use the following name convention for snapshots: muram\_en\_iterationnumber\_time  where ‘‘en'' stands for enhanced network and the time is measured in sec after switching on the \ac{ne} computation in the code, see also \citetalias{2024A&A...692A...6O}.}, we show additionally the \ac{sim} and \ac{cm} computations. We compare the spectra to an average \ac{qs} observation from the \ac{ftsatlas} \citep{1984SoPh...90..205N}. We begin by discussing the line profiles in comparison to the observation. After this, we discuss the asymmetry of the line profiles.

In Fig.~\ref{fig:fig1_av_spectra_and_bisector} (a), we show the \ac{qs} disk center \ac{ftsatlas} spectrum together with synthetic computations in a large wavelength window around the line center. The temporal average consists of four spatially averaged spectra from snapshots separated by approximately $2 \min$ of simulation time. The spatially averaged line profiles of these snapshots are shown in App.~\ref{app:time-variation}. In the far wings (\ie, $|\lambda - \lambda_{\mathrm{^{40}Ca,rest}}| > 2.5 \, \AA$), which form at the lower photosphere, the synthetic intensity approximately matches the observed spectrum and the differences between the computations are small. The equivalent widths of the synthetic spatially averaged profiles are $2.76\,\AA$ (\ac{im}), $2.74\, \AA$ (\ac{sim}), and $2.71\,\AA$ (\ac{cm}), and thus are rather similar. The slightly higher synthetic intensity compared to the observation might be due to the absence of blend lines in the \ac{rt} computation. For a more precise analysis of the line core, we present in panel (b) a zoom-in (as indicated by the dashed box in panel a). Here, the differences between the \ac{im}, \ac{cm}, and \ac{sim} computations from snapshot \snapfiveOneEight become more visible. We determined the line widths of the line core (as described in App.~\ref{sec:appendix_determination_of_line_width}) of the plotted profiles which are  $0.47 \, \AA$ (\ac{ftsatlas}), $0.48\, \AA$ (\ac{im}), $0.49 \,\AA$ (\ac{cm}), and $0.46\,\AA$ (\ac{sim}). Thus, the line width of the \ac{im} and \ac{cm} computations are close to the observed profile, and even $2 \,\%$ (\ac{im}) and $4\,\%$ (\ac{cm}) larger. The computation with only the most abundant isotope (\ac{sim}) results in an approximately $2 \%$ smaller line width compared to the \ac{im} computation. The line width of the temporally averaged snapshots is $0.48\, \AA$, and hence there is no significant difference between the time-averaged spectrum (red color) and that from a single snapshot (orange color). We note, however, that more snapshots are needed with a higher time resolution for a better comparison, which comes with additional computational costs. 

We find that the line profile minima of all computed spatially averaged line profiles are redshifted with respect to the rest wavelength of the most abundant calcium isotope $\lambda_{\mathrm{^{40}Ca, rest}}$. In the forward modeled spectra, the vertical component of the velocity at the formation height of the line profile minima is strongly correlated with the resulting Doppler shift of the line profile minima. Due to the imbalance of upflows ($43\%$) and downflows ($57\%$) at the formation heights of the line profile minima (see Sect.~\ref{sec:results_intensity_map_and_atmosphere} and Fig.~\ref{fig:fig2_int_vz_bz_temp} e), the minimum of the spatially averaged spectral line is redshifted. A similar result was found by \citet{2013SoPh..288...89C} who analyzed observations that suggest that the \caefft line forms slightly preferred in downflows. While the redshift of the \ac{im} computation from \snapfiveOneEight seems to be in good agreement with the observation (see Fig.~\ref{fig:fig1_av_spectra_and_bisector} d), we would like to note that the line shift varies by a small amount of $\approx 16 \, \mathrm{m\AA}$ with time. This difference can be seen in Appendix~\ref{app:time-variation}.

We now compare the line width from the spatially averaged line profile from snapshot \snapfiveOneEight with the averaged line width of single profiles. To this end, we calculated the line width for the \ac{im} computation from single profiles and found an average value of $0.37 \, \AA$, which is approximately $23 \, \%$ lower than what we found for the line width of the spatially averaged profile ($0.48\, \AA$, \ac{im}). We checked to what extent the line width of the spatially averaged line profile is due to the Doppler shifts of single profiles. We therefore calculated, similar to \citet{2009ApJ...694L.128L}, the standard deviation of the Doppler shift of the line profile minima $\sigma_{\Delta v, \mathrm{min}}$ from all spectra. We obtained a value of $\sigma_{\Delta v, \mathrm{min}}=4.56 \kms$, which might explain the larger line width of the spatially averaged spectrum\footnote{As a rough approximation, we assume Gaussian line shapes. Then $\sigma_{\Delta v,\mathrm{min}}=4.56\,\kms$ translates to a \ac{fwhm} of $0.31\, \AA$ at $\lambda=8542\,\AA$ and thus a convolution leads to a combined FWHM of $\sqrt{\left(0.31\,\AA\right)^2+\left(0.37\,\AA\right)^2}=0.48\,\AA$.}. 

This value is approximately a factor of four times higher than what \cite{2009ApJ...694L.128L} found in their forward modeled spectra and even higher than what these authors found from their observational data taken by the CRISP instrument \citep{2008ApJ...689L..69S} at the Swedish $1 \, \mathrm{m}$ Solar telescope. This suggests that in the \acs{muramche} model, the line width of the spatially averaged line profile partly results from single lines that are Doppler-shifted by the dynamic velocities in the simulated chromosphere.

While the line width of the \ac{cm} computation is larger than in the \ac{im} computation, the line profile of the \ac{cm} computation is slightly narrower near the minimum intensity. The location of this difference in the line profile coincides with the rest wavelengths of the less abundant calcium isotopes \citep[see \eg,][Fig.~1]{2014ApJ...784L..17L}. This hints that the fundamental assumption used to construct the composite model atom breaks down in the simulated chromosphere, which is indeed the case as we show in App.~\ref{app:composite_model_atom}.

\subsection{Asymmetry of spatially averaged line profiles}
\label{sec:results_average_profiles_bisectors}

The asymmetry of a spectral line can be visualized by studying its bisector. In its simplest way, the bisector of a spectral line can be constructed via 
\begin{align}
    b(I) = \frac{1}{2}(\lambda_{\mathrm{red}}(I) + \lambda_{\mathrm{blue}}(I)).
    \label{ref:eq_bisector}
\end{align}
\noindent Here, $I$ is the intensity, $\lambda_{\mathrm{red}}$ and $\lambda_{\mathrm{blue}}$ are the wavelengths on the red or blue side of the spectrum with respect to the line profile minimum $\lambda_{\mathrm{min}}$. In the literature, the bisectors of the \caefft line are shown at different reference wavelengths. They can be shown on an absolute wavelength scale \citep[see \eg,][]{2006ApJ...639..516U}, they can be centered on the bisector intensity in the far wings \citep[see \eg,][]{2013SoPh..288...89C}, or they can be shown centered on the wavelength of the line profile minimum $\lambda_{\mathrm{min}}$ as in \citet{2014ApJ...784L..17L}.

The bisector centered on an absolute wavelength scale, or centered on the far wing, allows for comparisons of line Doppler shifts in the line core.
The bisector as defined in \citet{2014ApJ...784L..17L} is given in terms of the wavelength relative to $\lambda_{\mathrm{min}}$ and not on the absolute wavelength scale. This allows for comparison of the shape and amplitude of different bisectors irrespective of the absolute wavelength calibration of the observation. For compatibility, we show our results in all three representations in Fig.~\ref{fig:fig1_av_spectra_and_bisector}. As an additional measure of line asymmetry, we determine for each bisector the amplitude $a_{\mathrm{core}}$, which is the distance in wavelength between $\lambda_{\mathrm{min}}$ and the red-most excursion. We call the intensity of the line profile at which the maximum red excursion is reached $I_{\mathrm{bs}}$. The amplitude $a_{\mathrm{wing}}$ is the distance between the red-most excursion and the bisector at $I_{\mathrm{c}}/2$, where $I_{\mathrm{c}}$ is the continuum intensity. For an overview of these quantities, see App. \ref{sec:appendix_determination_of_line_width} and Fig.~\ref{fig:fig_appendix_line_bisector}.

In Fig.~\ref{fig:fig1_av_spectra_and_bisector} (panels c, d, and e), the line bisectors of the spatially averaged profiles shown in panel (b) are presented. The \ac{ftsatlas} line profile shows the typical ‘‘inverse-C shaped'' line bisector with an amplitude of $a_{\mathrm{core}} = 14.98 \, \mathrm{m\,\AA}$ at an intensity of $I_{\mathrm{bs}} = 11.86 \, \intensity$, the corresponding line wing amplitude of the bisector is $a_{\mathrm{wing}}=34.47 \, \mathrm{m\,\AA}$. The line profiles computed with the \ac{im} or \ac{cm} show amplitudes of $a_{\mathrm{core}} = 19.44 \, \mathrm{m\,\AA}$ (\ac{im}) and $a_{\mathrm{core}} = 14.74 \, \mathrm{m\,\AA}$ (\ac{cm}) at intensities of $I_{\mathrm{bs}} = 10.98$~(\ac{im}) and $I_{\mathrm{bs}} = 11.83$ (\ac{cm}). Thus, in this particular snapshot, the amplitude and intensity of the \ac{cm} computation match the observed bisector better than the full isotope computation (\ac{im}). We note, however, the minimum intensity of the line profile in the \ac{cm} computation is higher than in the \ac{im} computation, which explains the shift of the bisector intensity. The wing amplitudes of the \ac{im} and \ac{cm} computations are $a_{\mathrm{wing}} = 56.52 \, \mathrm{m\,\AA}$ (\ac{im}) and  $a_{\mathrm{wing}} = 45.16 \, \mathrm{m\,\AA}$~(\ac{cm}) which are both larger than in the observation. By taking only the most abundant isotope of calcium into account, the bisector of the \ac{sim} computation does not show a visible red asymmetry, and hence no clear ‘‘inverse C-shape''. The bisector from the time-averaged \ac{im} computations results in an $a_{\mathrm{core}} = 16.16 \, \mathrm{m\,\AA}$, which is closer to the observed value and indicates that the amplitude is time-dependent. The wing amplitude of the time-averaged spectra is the largest with a value of $a_{\mathrm{wing}}=61.19\, \mathrm{m\,\AA}$, displaying that even in the wings of the lines, the average spectrum is sensitive to time averaging. 

In panel (d), the bisectors are shown on an absolute wavelength scale. The wavelength axis is centered on the rest wavelength of $^{40}\mathrm{Ca}$ to avoid large wavelength numbers in the plot. It can be seen that above $20 \intensity$ the computations accounting for isotopic splitting agree with each other. The \ac{sim} computation is different, which suggests that the isotopic splitting effect is still visible outside the line core.

In Fig.~\ref{fig:fig1_av_spectra_and_bisector} panel (e) the bisectors are shown centered on the wavelength where the bisector intensity is $60 \%$ of the continuum intensity, which is similar to the representation in \citet{2013SoPh..288...89C}. In this representation, the maximum redshift of the \ac{im} bisector is $61.8 \, \mathrm{m\,\AA}$, which is larger than the $34.3 \, \mathrm{m\,\AA}$ for the \ac{sim}  computation, highlighting again the isotopic splitting effect on the asymmetry of the spatially averaged line profile. The \ac{im} computation compares approximately to the value found by \citet{2013SoPh..288...89C}, 
which was $\approx 51 \mathrm{\,m\AA}$ ($5.1\,\mathrm{pm}$) in their quiet region and $\approx58 \mathrm{\,m\AA}$ ($5.8\,\mathrm{pm}$) in their
internetwork region. We note that we have not degraded
our spectra with an instrumental PSF. In addition, the observed region of \citet{2013SoPh..288...89C}
might contain a different amount of magnetic flux than our simulation.

\subsection{Dependence of line profiles on atmospheric conditions}
\label{sec:results_average_spectra_depending_on_atmosphere}
\begin{figure*}
   \centering
   \sidecaption
   \includegraphics[width=12cm,clip]{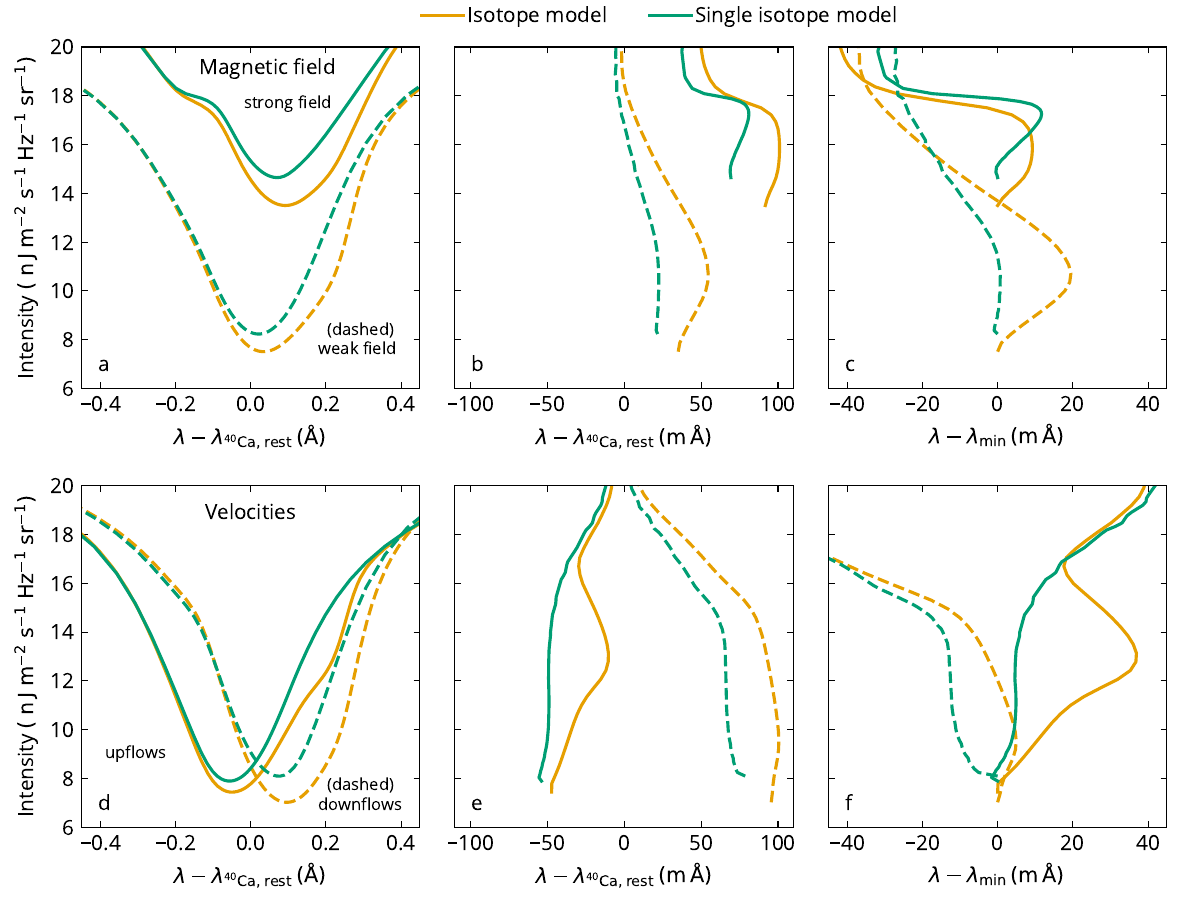}
      \caption{Profiles and bisectors averaged over pixels with similar atmospheric properties. We show data from synthetic spectra once taking all isotopes into account (orange) and once taking only the most abundant isotope (green) into account. The top row (panels a, b, and c) shows results computed from a selection of strong (solid lines) and weak (dashed lines) magnetic fields in the formation height region of the line core (see text for details). The bottom row (panels d, e, and f) shows only profiles and bisectors over the pixels harbouring upflows (solid lines) or downflows (dashed lines) in the formation height region of the line core (see text). The bisectors are shown once on an absolute wavelength scale (panels b and e) and on a scale where the bisectors are centered on the line profile minimum wavelength $\lambda_{\mathrm{min}}$. The synthetic data corresponds to snapshot \snapfiveOneEight.}
         \label{fig:av-spectra-and-bisectors-depending-on-vz-and-bz}
\end{figure*}

In Sects.~\ref{sec:results_average_profiles} and \ref{sec:results_average_profiles_bisectors}, we discussed the shape and asymmetry of the synthetic line profiles that were averaged over the whole computational domain in comparison with the observation. There are two mechanisms that contribute to the width and asymmetry of the line, namely the atmospheric structure and the presence of multiple calcium isotopes. In this section, we study how the atmospheric structure affects the line profiles.
We study the shape and asymmetry of the \caefft line by preselecting columns based on their atmospheric properties before computing the averages. We distinguish between columns based on their velocity structure and magnetic field strength along the \ac{los}. For each vertical column in the atmosphere, we computed the  average vertical velocity $v_{\mathrm{avg}}$ between the formation height of the inner wing intensity and the line core, such that:
\begin{align}
    v_{\mathrm{avg}} = \frac{\int_{z1}^{z2} v_z (z') \,\mathrm{d} z'}{z_2-z_1}
    \label{eq:average_velocity_formation_region}
\end{align}
where
\begin{align}
    z_1 = \frac{z(\tau_{\mathrm{wing,v}}=1) + z(\tau_{\mathrm{wing,r}}=1)}{2}
\end{align}
and
\begin{align}
    z_2 = \max_{\{\lambda\; | \; ||\lambda - \lambda_0|| = 1 \AA\}}z(\tau_{\lambda}=1).
\end{align}

Here $v_z$ is the vertical velocity in the atmosphere, $z_1$ is the average formation height of the wing intensity at $\pm 1 \AA$ from the rest wavelength $\lambda_0$ of \caisotopefourty, and $z_2$ is the maximum formation height in the wavelength window $\pm 1 \AA$ around the rest wavelength. We compute the averaged unsigned vertical magnetic field $|B_z|_{\mathrm{avg}}$ analogously to the average vertical velocity. 

We categorize columns with  $|B_z|_{\mathrm{avg}}<60 \G$ as having weak \ac{los} magnetic field and $|B_z|_{\mathrm{avg}}>60 \G$ as having strong \ac{los} magnetic field. This separation roughly distinguishes between columns outside and inside the large-scale network field (see Fig.~\ref{fig:fig2_int_vz_bz_temp}, panel f). Similarly, we distinguish between columns having $v_{z,\mathrm{avg}}>0$ as upflows and $v_{z,\mathrm{avg}}<0$ as downflows. The different categorizations between vertical motions and magnetic fields are not meant to form stochastically independent sets.

In Fig.~\ref{fig:av-spectra-and-bisectors-depending-on-vz-and-bz} we show how the average profiles and bisector shapes vary depending on magnetic field (panels a, b, and c) and vertical velocity (panels d, e, and f). The bisectors are presented once relative to the rest wavelength of $^{40}$Ca (panels b and e) and once relative to the line profile minimum (panels c and f).
In Fig.~\ref{fig:av-spectra-and-bisectors-depending-on-vz-and-bz} (a), it can be seen that the line profiles of the \ac{wfp} all appear similar to their corresponding profiles in Fig.~\ref{fig:fig1_av_spectra_and_bisector}. This is expected because approximately $ 93 \%$ of all profiles in this snapshot are formed in the weak field regions. The \ac{sfp}, which are averaged over approximately $7 \%$ of all spectra, have higher intensities across the whole wavelength range shown. The continuum at $10 \, \AA$ away from the line core is at roughly $41 \intensity$, which is comparable to the \ac{wfp}. In App.~\ref{app:line_with_continuum} we show the line profiles of the \ac{im} computation over a larger wavelength window. The corresponding bisectors of the \ac{sfp} in Fig.~\ref{fig:av-spectra-and-bisectors-depending-on-vz-and-bz} (b and c, solid lines) all show an inverse ''C‘‘ shape and a blueward turn close to the wing intensity, irrespective of the different treatments of isotopic splitting in the \ac{rt} computation.

In the bottom panels (d), (e), and (f), we show the resulting \ac{ufp} in solid lines and \ac{dfp} in dashed lines, see Eq.~(\ref{eq:average_velocity_formation_region}). In this snapshot, we find $45.5\%$ of the profiles form in an upflow and $54.5\%$ form in a downflow. These numbers compare approximately to the vertical velocity at the formation height of the line profile minimum (see Sect.~\ref{sec:results_intensity_map_and_atmosphere} and Fig.~\ref{fig:fig2_int_vz_bz_temp}~e).

In Fig.~\ref{fig:av-spectra-and-bisectors-depending-on-vz-and-bz}~(d), it can be seen that \ac{ufp}, besides being blue-shifted, have slightly higher intensities in the minimum. The differences in the \ac{ufp} between the  \ac{im} and \ac{sim} \ac{rt} computations are negligible bluewards of the line core, but there are significant differences redwards. The corresponding bisectors of the \ac{ufp} (panels e and f, solid lines) are blue shifted (panel e) and show a red asymmetry close to the line core with respect to the line profile minimum (panel f). Even for the \ac{sim} (green curve), a slight red asymmetry in the line core is visible. In the outer wings, that is for intensities above $\approx 18 \intensity$, the averaged \ac{ufp} become approximately symmetric around the rest wavelength of $^{40}$Ca (panel e).

Unsurprisingly, the \ac{dfp} are red-shifted with respect to the rest wavelength of $^{40}$Ca. Similar to the \ac{ufp}, in the \ac{dfp}, the different \ac{rt} computations are in good agreement bluewards to the line center and differ redwards. The corresponding bisectors (dashed lines) are redshifted (panel e), and indicate a much lower red asymmetry with respect to the line profile minimum (panel f). While the \ac{im} computation shows a much lower bisector amplitude compared to the spatially averaged line profile computed over the whole simulation domain (see Fig.~\ref{fig:fig1_av_spectra_and_bisector} c), the bisector from the \ac{sim} remains blueshifted at all intensities with respect to the line profile minimum.

This study can be summarised as follows: The strengths and shapes of the line profiles and their corresponding bisectors can depend strongly on regions selected by the \ac{los} velocity or the \ac{los} magnetic field strength in the atmosphere. In strong \ac{los} magnetic field regions, even the averaged profile of the computation with only the most abundant isotope (\ac{sim}) shows an inverse ''C‘‘ shape. But in regions of weaker \ac{los} magnetic field or downflows, the \ac{sim} computation does not result in an inverse C-shaped line bisector, that is, when isotopic splitting is not taken into account.

\section{Summary and discussion} 
\label{sec:discussion}

In this work, we presented spatially averaged synthetic spectra of the \caefft line, computed from an \acs{muramche} \ac{rmhd} model, which we compared to an average \ac{qs} observation from the \ac{ftsatlas}. Including a large-scale bipolar magnetic feature, the atmosphere model resembles an enhanced network region. We computed three different sets of synthetic spectra in 1.5D \ac{rt}. We calculated the spectra once taking all abundant isotopes of calcium in the solar atmosphere into account, with the transitions due to the various isotopes treated as blend lines (\ac{im}), and once where only the most abundant isotope was considered (\ac{sim}). In addition, we utilized an approximate model of isotopic splitting that combines the effects of isotopes in only one model atom (\ac{cm}).

The two-dimensional image of the line-core intensity (obtained from the \ac{im} computation) shows a typical shock-expansion pattern \citep[see also][]{2004A&A...414.1121W} in the quiet part of the simulation. Above the network magnetic field, the intensity is enhanced. In the intensity image, no fibrilar structures are seen, while they are visible in the formation height of the minimum intensity. This is likely attributed to the lack of horizontal \ac{rt} similar to the H$\alpha$ line \citep{2012ApJ...749..136L}. The formation height of the line core shows large spatial variation with fine-structured details. 

The vertical component of the velocity at the formation height of the line center shows thinner upflow structures that are surrounded by expanded downflow structures. This suggests upward mass flows in thin dense channels, while the mass falls back in blobs of lower density, which cover a larger volume in the atmosphere.

We then studied the spatially and temporally averaged profiles for the \ac{im} computation and spatially averaged profiles from one snapshot for the \ac{im}, \ac{sim}, and \ac{cm} computation. We find a good match between the spatially and temporally averaged \ac{im} computation and the observation in terms of the line profile depth, width, and shape, including the asymmetry as expressed by the line bisector. When only the most abundant isotope of calcium is considered in the synthesis, the resulting average spectrum is slightly narrower and almost symmetric. These findings confirm the results of \citet{2014ApJ...784L..17L}, who suggested that isotopes must be taken into account to reproduce the observed inverse C-shape bisector. The CM calculation similarly reproduces the overall line width and asymmetry of the IM calculation. In the innermost line core, the intensity in the \ac{cm} computation is, however, slightly higher redwards of the line profile minimum. In the composite model approach, it is assumed that the ratios of population-level densities from different isotopes are constant throughout the whole atmosphere. In App.~\ref{app:composite_model_atom} we show that this approximation breaks down already in the lower to mid chromosphere where the \caefft line forms.

We found that the line width of single profiles is, on average, approximately $23\%$ smaller than the width of the spatially averaged line profile. Similar to \citet{2009ApJ...694L.128L}, we measured the standard deviation of the Doppler shifts of the intensity minimum of the line profiles. We found a value of $\sigma_{\Delta v, \mathrm{min}}=4.56 \kms$, which might explain the difference between the line widths of single profiles and the width of the spatially averaged line profile. The obtained $\sigma_{\Delta v, \mathrm{min}}$ is approximately four times higher than in \citet{2009ApJ...694L.128L}. This suggests the close match of the line width in the \acs{muramche} simulation with the observation is due to the dynamic atmosphere. \citet{2009ApJ...694L.128L} compared their results with observations from a coronal hole, which showed a value of $\sigma_{\Delta v, \mathrm{min}}=2.2 \kms$. The reason for the roughly two times higher value we find in the \acs{muramche} simulation could be similar to the result of \citet{2018ApJ...864...21K} who found a larger line width of the \mgk line in the \ac{qs} compared to a coronal hole. This suggests the simulation presented here compares well to the \ac{qs} but a different simulation might be needed for the comparison with coronal hole observations. The observations from the third flight of the Sunrise balloon-borne observatory \citep[Sunrise III,][]{2025SoPh..300...75K} will provide superior seeing-free observations of the \caefft line arising from different regions on the Sun. A comparison between these observations and the computations presented here and future \acs{muramche} simulations will help to better understand the dynamics of the chromosphere.

We also considered how the shape of the bisector not only depends on whether all isotopes of calcium are included but can also be strongly influenced by atmospheric conditions. To this end, we separated the atmosphere into four sets of columns with strong or weak vertical magnetic fields and up- or downflows measured over the formation height range of \caefft. These sets are not disjoint but still show different features in the average spectra. We found that average spectra over columns where $|B_{z}|_{\mathrm{avg}} < 60 \G$ ($93\%$ of the spectra) match the observed \ac{ftsatlas} spectra best in terms of the line shape and the amplitude of the bisector. When the average is taken over profiles from columns where $|B_{z}|_{\mathrm{avg}} \geq 60 \G$, the line intensity is weakened, and even the \ac{sim} computation shows an ‘‘inverse C shape'' which is comparable in amplitude to the other synthetic computations. The higher intensity is a result of the higher temperature in the regions dominated by a stronger magnetic field. The asymmetry must be a result of the flow structure above the magnetic features. 

The separation into up- and downflows results in multiple effects. The asymmetries of the line shapes are substantially different, as measured by the line bisectors. The averaged profiles from columns showing downflows are much less asymmetric at low intensities, that is, close to the line core. The profiles averaged over regions showing on average upflows show a red asymmetry close to the line core. 

The comparison presented here between different atmospheric parameters and the resulting spectral line properties is intended to understand the effect of isotopic splitting at different locations in the simulation domain in an average sense. \citet{2024A&A...682A..11M} demonstrated that line profiles from single resolution elements in a simulation can be more complicated than individual observed profiles. This trend even holds after the synthetic profiles were degraded to match the instrumental conditions. In a follow-up study, we plan to do a detailed comparison between the simulation and high-resolution observations.

\section{Conclusions} 
\label{sec:conclusions}

We could show that the \acs{muramche} code can produce an atmosphere model that results in a close match with observations. For this purpose, we forward modeled the chromospheric  \caefft spectral line and found a good agreement in comparison with observations from the \ac{ftsatlas}. The reasonably good match of the line width is a result of the dynamic velocity field at chromospheric heights in the \acs{muramche} model. The line profile minimum is shifted towards the red with respect to the rest wavelength and approximately matches the position of the observed \ac{ftsatlas} line profile. In the \acs{muramche} model, the net shift is a result of an imbalance between up and downflows in the chromosphere at the formation height of the \caefft line core. Whether a similar trend is apparent in high-resolution observations will be investigated in a follow-up study with Sunrise III observations.

We confirmed the need for multiple calcium isotopes in the forward modeling to reproduce the observed inverse C-shape asymmetry of the line profile. The asymmetry in the spectral profile produced by isotopes can, however, be mimicked by other atmospheric properties such as the magnetic field and the velocity structure. As a consequence, if the effects of multiple isotopes are not taken into account during the inversion of the \caefft line profile then the inversion code will likely compensate for the missing isotopes by tweaking velocity and/or magnetic field structures and thus introducing errors in the inferred atmosphere. 

Overall, we conclude that the model atmosphere must be sufficiently dynamic and the isotopic splitting effect must be taken into account in the forward modeling to match the observed \caefft line profile.

\begin{acknowledgements}

 We thank the anonymous referee for comments and suggestions that improved the quality of this paper. P.O. would like to thank J. Leenaarts and J. de la Cruz Rodriguez for discussions about the isotopic splitting effect. In addition, P.O. would like to thank T. M. D. Pereira for support in using the RH1.5D code. This project has received funding from the European Research Council (ERC) under the European Union’s Horizon 2020 research and innovation programme (grant agreement No. 101097844 — project WINSUN). This work was supported by the International Max-Planck Research School (IMPRS) for Solar System Science at the University of Göttingen. This work was supported by the Deutsches Zentrum f{\"u}r Luft und Raumfahrt (DLR; German Aerospace Center) by grant DLR-FKZ 50OU2201. We gratefully acknowledge the computational resources provided by the Cobra and Raven supercomputer systems of the Max Planck Computing and Data Facility (MPCDF) in Garching, Germany. D.P. would like to thank A. Irwin (Free-EoS).

\end{acknowledgements}

\bibliographystyle{aa}

\bibliography{bib-ads}

\begin{appendix}

\section{Time variation of spatially averaged line profile}
\label{app:time-variation}
\begin{figure}
   \centering
   \includegraphics[width=\hsize,clip]{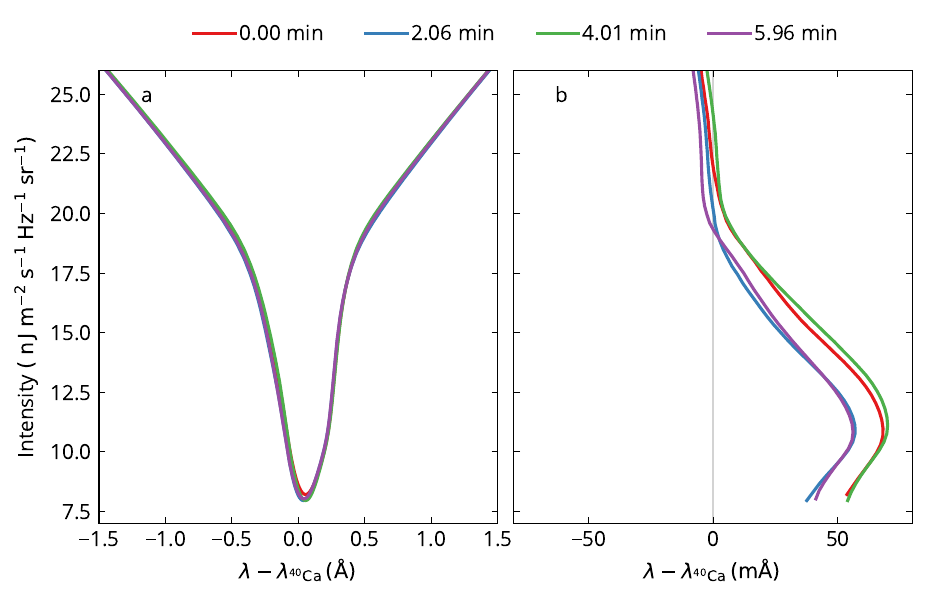}
      \caption{Time variation of spatially averaged line profile in the simulation. We show the spatially averaged line profiles (panel a) and their bisectors (panel b) from four snapshots of the simulation that are averaged by approximately $2 \min$ of simulation time. The snapshot at $t=2.06 \min$ (blue curves) corresponds to the snapshot \snapfiveOneEight that was discussed in detail in the main text. The line profiles are shown in a large wavelength window to show that there is also a time variation in the far wings of the spectral line in the simulation. The bisectors in panel (b) are shown on an absolute wavelength scale, similar to Fig.~\ref{fig:fig1_av_spectra_and_bisector} (d). All the synthetic profiles in this figure were computed including isotopes and thus correspond to \ac{im} computations.
      }
         \label{fig:fig_appendix_time_variation}
\end{figure}

The \ac{ftsatlas} observation represents an average region of the quiet Sun that is averaged over space and time. In the main text in Sect.~\ref{sec:results_average_profiles} we discussed the spatially averaged line profile from one snapshot of the simulation, and a profile that was averaged over four snapshots. To provide more context on the temporal variation of the spatially averaged line profiles from these four snapshots, we present them here explicitly in Fig.~\ref{fig:fig_appendix_time_variation} on an absolute wavelength scale. The four snapshots are separated by approximately $2 \min$ of simulation time. The snapshot at $2.06 \min$ corresponds to snapshot \snapfiveOneEight, which is discussed in detail in the main text. The red curve in Fig.~\ref{fig:fig1_av_spectra_and_bisector} corresponds to the temporal average of the four snapshots shown in Fig.~\ref{fig:fig_appendix_time_variation}. It can be seen that even outside the line core, which is roughly above $20 \intensity$, the bisector shows time variations. This suggests that the simulated lower atmosphere displays oscillations. Using the bisector of these far wings as a reference might thus be time-dependent. For a reasonable estimate of such a reference wavelength, many more synthesized snapshots are required.

\section{Determination of line width and bisector parameters}
\label{sec:appendix_determination_of_line_width}

\begin{figure}
   \centering
   \includegraphics[width=\hsize,clip]{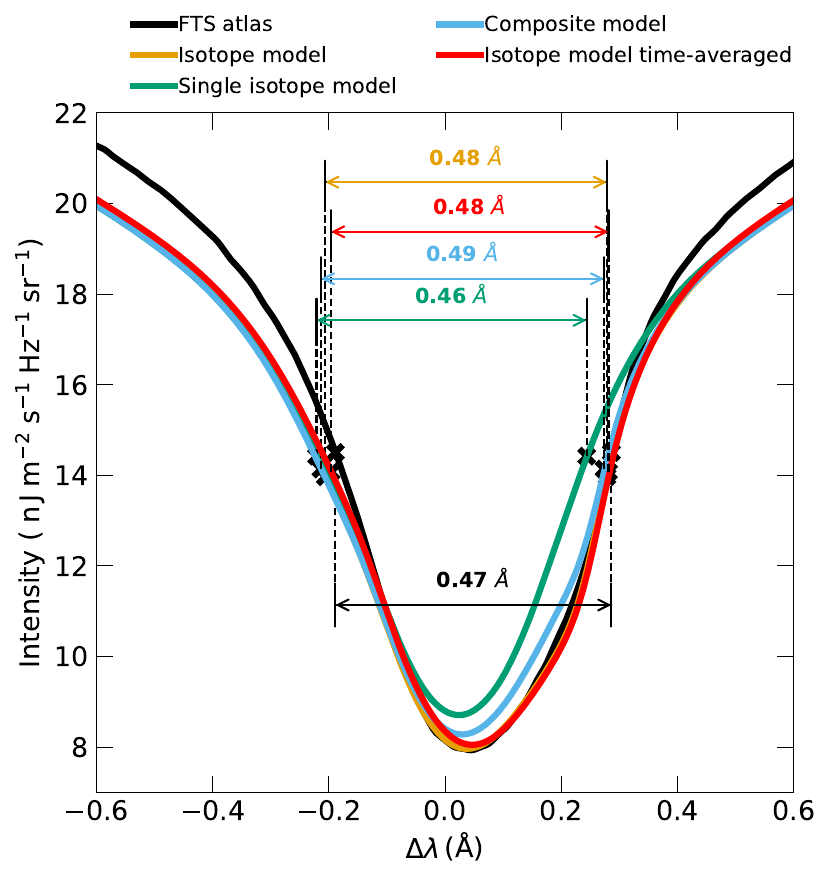}
      \caption{Determination of line width. The line width is measured at half the intensity between the wing intensity $I_{\mathrm{wing}}$ and the profile minimum intensity $I_{\mathrm{min}}$. The intensity $I_{\mathrm{wing}}$ is taken as the average intensity at $\pm 0.6 \AA$ away from the line profile minimum. The obtained line widths for the different computations and the observation are indicated in the figure.
      }
         \label{fig:fig_appendix_line_width}
\end{figure}

We determine the line width following the description in \citet{2009A&A...503..577C}. First, we determine the wavelength position of the profile minimum $\lambda_\mathrm{min}$. From that wavelength position, we take the average intensity in the line wings at the positions $\lambda_{\mathrm{wing}}$ that fulfill $|\lambda - \lambda_{\mathrm{min}}| = 0.6 \AA$. The line width is then given by the wavelength difference of the intensity profile at half the intensity between the minimum and the wing intensity. This technique was used instead of just the \ac{fwhm} to make sure that we are considering the width of the line core, which is the important part of the line for the chromosphere. In this way, we avoid the influence of the prominent line wings. The measured line widths are indicated in Fig.~\ref{fig:fig_appendix_line_width}.

\begin{figure}
   \centering
   \includegraphics[width=\hsize,clip]{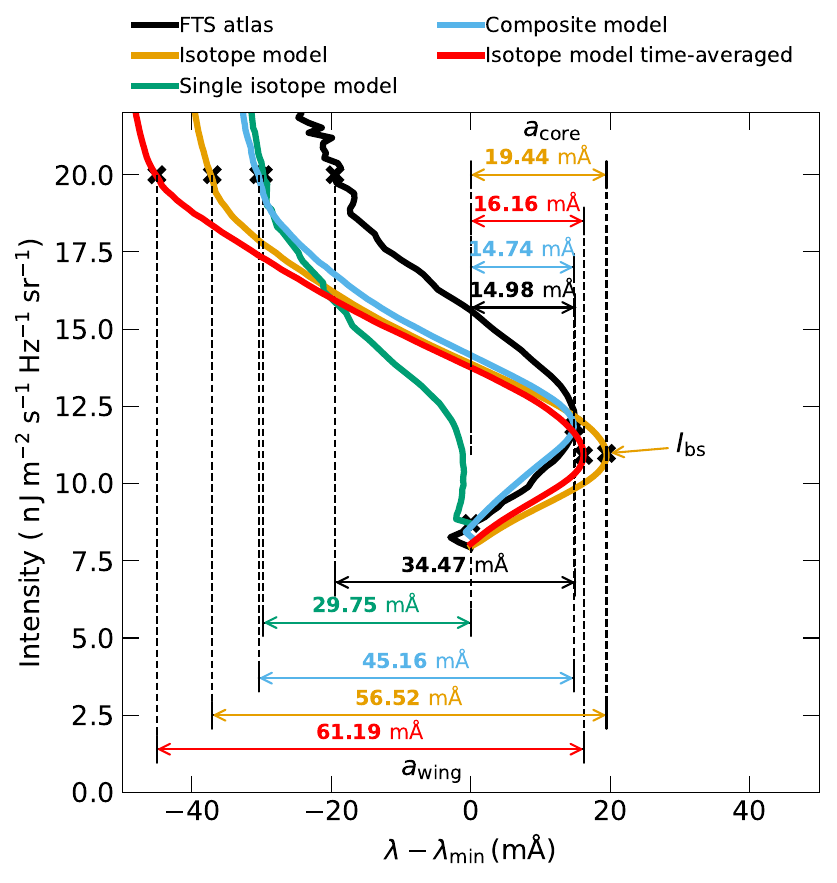}
      \caption{Determination of line bisector parameters. Shown are the line bisectors for the spatially averaged line profiles for the observation (black), the \ac{im} computation (orange), the \ac{cm} computation (blue), the \ac{sim} computation (green), and the temporal average of four snapshots computed with the \ac{im} treatment (red). The core amplitude of the bisector $a_{\mathrm{core}}$ is the wavelength difference between the line profile minimum and the red-most excursion of the bisector. The wing amplitude is the wavelength difference between the red-most excursion and the bisector at approximately half the continuum intensity.
      }
         \label{fig:fig_appendix_line_bisector}
\end{figure}

We describe the bisector by parameters similar to those used in \citet{2011ApJ...736..114P} and \citet{2013ApJ...764..153P}. The bisector amplitude in the line core $a_{\mathrm{core}}$ is defined as the difference in wavelength between the profile minimum $\lambda_\mathrm{min}$ and the red-most excursion of the bisector. The intensity at the red-most excursion is named $I_{\mathrm{bs}}$. The wing amplitude $a_{\mathrm{wing}}$ of the bisector is defined as the difference between the red-most excursion of the bisector and the bisector at an intensity of $\approx I_{\mathrm{c}}/2$ where $I_{\mathrm{c}}$ is the continuum intensity at $10 \, \AA$ away from the profile minimum. An overview of the measured bisector amplitudes is shown in Fig.~\ref{fig:fig_appendix_line_bisector}.

\section{The composite model atom}
\label{app:composite_model_atom}

Here, we briefly describe that the fundamental assumption behind the composite model is, in general, not fulfilled in the simulated chromosphere.

\label{sec:composite_model_atom_population_level_ratios}
In the composite model, the absorption coefficient is constructed by summing up Voigt profiles centered at the corresponding rest wavelengths of the different isotopes. The sum is weighted by the relative abundance of the isotope with respect to the most abundant isotope $^{40}$Ca. This approach is an approximation because it is assumed that population-level ratios between different isotopes are constant throughout the line formation region \citep{ 1986UppOR..33.....C,2014ApJ...784L..17L}. \cite{2014ApJ...784L..17L} discussed that this is only an approximation, but can be justified by insignificant differences from the full computation. As shown in Fig.~\ref{fig:fig1_av_spectra_and_bisector}, we find in the \acs{muramche} model differences towards the red from the line profile minimum between the \ac{im} and \ac{cm} computation.

\begin{figure*}
   \sidecaption
\includegraphics[width=\textwidth,clip]{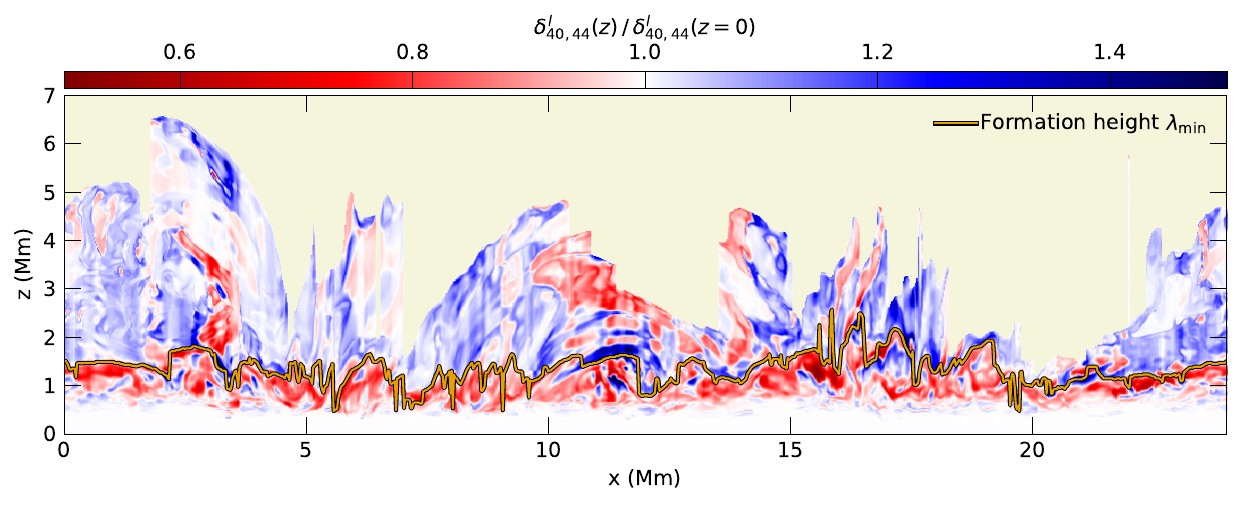}
      \caption{Validity of the composite atom model. We show the population level ratio of the lower energy state of the \caefft transition between $^{40}$Ca and $^{44}$Ca normalized by the ratio at $z=0$ (see text). $^{40}$Ca and $^{44}$Ca are the two most abundant Ca isotopes. The orange curve indicates the formation height of the intensity minimum of the \caefft line. The shown data is from the \ac{im} computation of snapshot\snapfourNineNine. The light-yellow background indicates where no data is plotted.}
         \label{fig:appendix_population_ratios}
\end{figure*}

In Fig.~\ref{fig:appendix_population_ratios}, we compare the population ratios in a cut through the atmosphere. We compute this ratio for the two most abundant isotopes, which are $^{40}\mathrm{Ca}$ and $^{44}$Ca, and show that the ratios of the populations in the atmosphere deviate with height. We computed the ratio of the population-level densities (here for the lower level ''l`` of the transition) via:
\begin{align}
\popratio{40}{44}{l}.
\end{align}
Here, $n$ is the population-level density of the respective calcium isotopes. The population-level ratio for the upper level (not shown) is computed analogously and shows a similar behavior. We normalized the height-dependent ratio by the ratio at $z=0$. In the lower atmosphere, at about $z\leq 0.5 \Mm$, \popratioshort{40}{44}{l} is roughly constant. Higher up in the atmosphere, but still below the formation height of the line profile minimum (as indicated by the orange curve), the \popratioshort{40}{44}{l} starts to deviate. The assumption on which the composite model is based breaks down at these heights in the simulated chromosphere.

\section{\caefft line with continuum}

\label{app:line_with_continuum}

\begin{figure*}
   \centering
   \includegraphics[width=\hsize,clip]{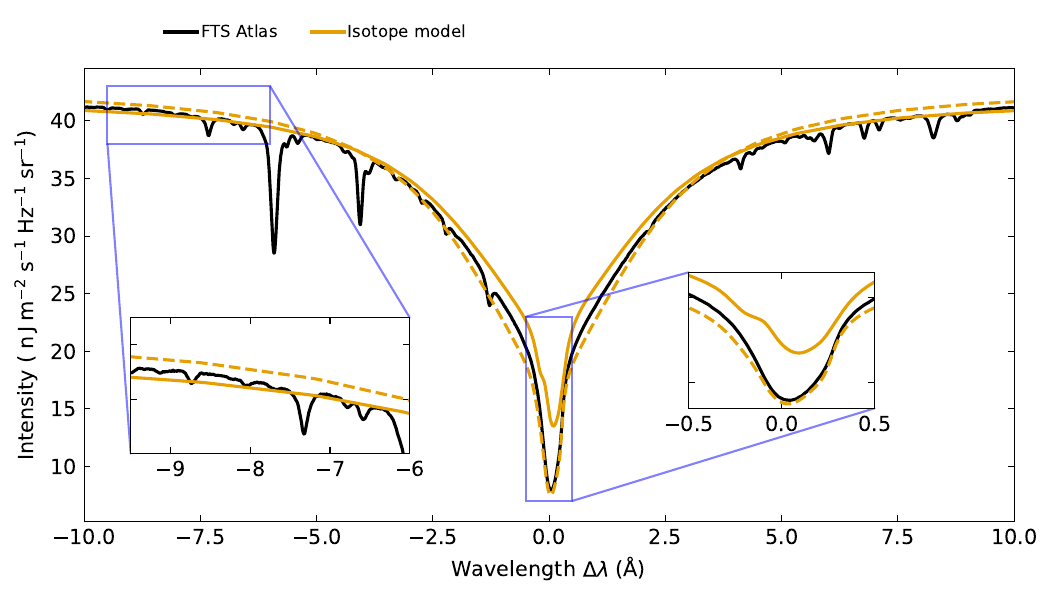}
      \caption{\caefft line averaged over regions of different vertical magnetic fields. The layout is similar to Fig.~\ref{fig:av-spectra-and-bisectors-depending-on-vz-and-bz} (panel a) but shows a larger wavelength range, which also includes the continuum. The black line shows the observed data from the \ac{ftsatlas}. The dashed orange line indicates the line profile of the \ac{im} computation averaged over regions which have $|B_z|_{\mathrm{avg}}<60 \G$ and the solid orange line that is averaged over regions which have $|B_z|_{\mathrm{avg}}\geq60 \G$ (see Sect.~\ref{sec:results_average_spectra_depending_on_atmosphere}). For simplicity, we show only the data for the \ac{im} computation. The inset panels show zooms to the continuum and the line core, the latter is similar to Fig.~\ref{fig:av-spectra-and-bisectors-depending-on-vz-and-bz} (panel a).
      }
         \label{fig:fig_appendix_full_line}
\end{figure*}

In the main text (Sect.~\ref{sec:results_average_spectra_depending_on_atmosphere}), we presented the intensity profile averaged over regions depending on the vertical component of the magnetic field averaged over the height of formation between the inner wings and the line core. Here, we present an overview of how the line profiles compare in a larger wavelength window containing the continuum. Figure~\ref{fig:fig_appendix_full_line} shows the intensity of the \ac{ftsatlas} observation and the averaged intensities over regions, which have $|B_z|_{\mathrm{avg}}<60 \G$ (dashed lines) those averaged over regions which have $|B_z|_{\mathrm{avg}}\geq60 \G$ (solid lines, see Sect.~\ref{sec:results_average_spectra_depending_on_atmosphere}). While the line core intensity of the \ac{sfp} profile is clearly enhanced, the difference between \ac{sfp} and \ac{wfp} profiles is small in the continuum.

\end{appendix}

\end{document}